\documentclass{article}
\usepackage{spconf,amsmath,graphicx}
\usepackage{subfigure}
\usepackage{amsfonts}
\usepackage{mathrsfs}
\usepackage{booktabs}
\usepackage{multirow}
\usepackage{caption} 
\title{A Multi-Stage Triple-Path Method for Speech Separation in Noisy and Reverberant Environments}
\name{Zhaoxi Mu$^{\star}$ \qquad Xinyu Yang$^{\star}$ \qquad Xiangyuan Yang$^{\star}$ \qquad Wenjing Zhu$^{\dagger}$}
  
  \address{$^{\star}$ School of Computer Science and Technology, Xi'an Jiaotong University, Xi'an, China \\
      $^{\dagger}$ Du Xiaoman, Beijing, China}

\begin{document}
%
\maketitle
\begin{abstract}
In noisy and reverberant environments, the performance of deep learning-based speech separation methods drops dramatically because previous methods are not designed and optimized for such situations. To address this issue, we propose a multi-stage end-to-end learning method that decouples the difficult speech separation problem in noisy and reverberant environments into three sub-problems: speech denoising, separation, and de-reverberation. The probability and speed of searching for the optimal solution of the speech separation model are improved by reducing the solution space. Moreover, since the channel information of the audio sequence in the time domain is crucial for speech separation, we propose a triple-path structure capable of modeling the channel dimension of audio sequences. Experimental results show that the proposed multi-stage triple-path method can improve the performance of speech separation models at the cost of little model parameter increment.
\end{abstract}
\section{Introduction}
Speech separation is a common speech preprocessing method that separates the clean speech of each source from the mixture. In this work, we focus on the problem of single-channel speech separation in noisy and reverberant environments. With the application and development of deep learning in speech separation, its performance has been greatly improved. For example, Luo and Mesgaran \cite{luo2019conv} proposed a time-domain speech separation method to avoid predicting phase independently. Luo et al. \cite{luo2020dual} proposed a dual-path method to process long speech sequences efficiently. On this basis, Subakan et al. \cite{subakan2021attention} and Chen et al. \cite{chen2020dual} use the powerful sequence modeling ability of Transformer to replace or combine with RNN to improve the performance further.

However, the collected speech is usually not just a mixture of the voices of multiple speakers but also contains various kinds of noise and reverberation in real acoustic environments. Noise masks the speech signal and corrupts the phase information. Reverberation causes spectral smearing of the source. Most speech separation methods suffer a sharp drop in performance in the face of noise and reverberation \cite{maciejewski2020whamr} because they are not designed and optimized for this situation. A simple solution is to cascade multiple models solving different problems and perform joint end-to-end optimization \cite{tan2018two,zhao2019two,delfarah2020talker,LiuDW20,tan2020audio}. However, compared with the single-stage method, the training process of the cascade method is cumbersome, and the memory consumption is large. Furthermore, when a scale-invariance loss function such as SI-SNR \cite{RouxWEH19} is used to train the cascade system, the output scales of each network may not match, resulting in severe performance degradation \cite{maciejewski2020whamr}. To address these issues, inspired by curriculum learning concept \cite{BengioLCW09,LiLLYZL21}, we design a single-model multi-stage training method, which can not only train an end-to-end network to simplify the training process but also outperforms single-stage methods with the same model size.

\begin{figure}[t]
    \centering
    \includegraphics[width=0.50\textwidth]{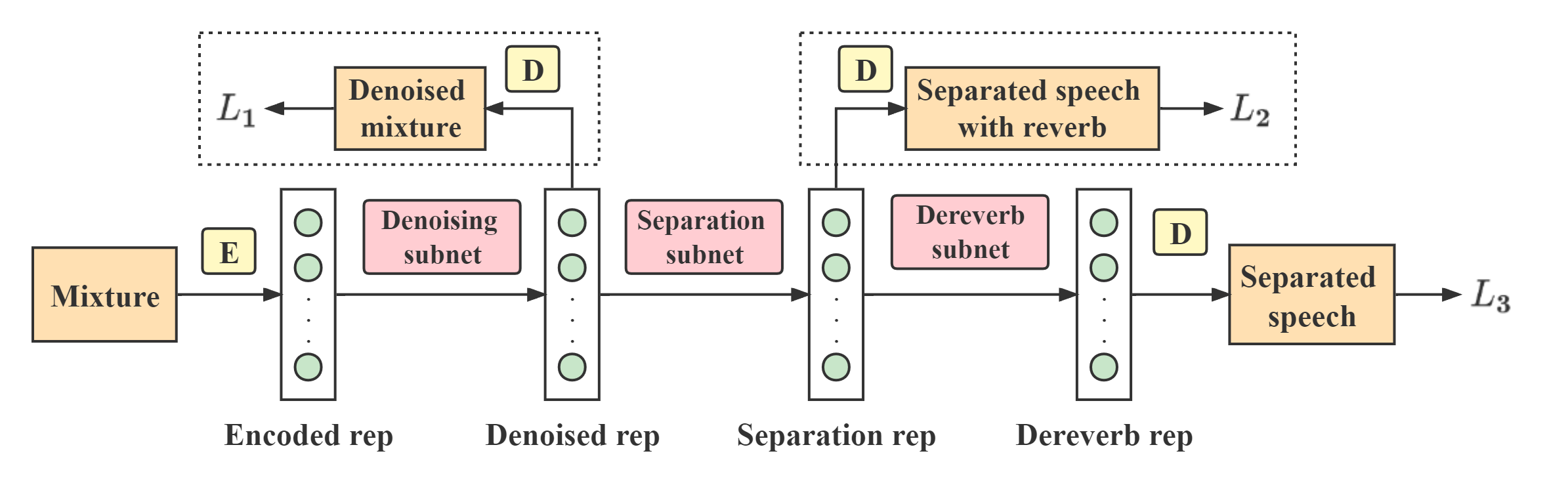}
    \caption{Schematic diagram of the multi-stage method. E, D, and rep are abbreviations for the encoder, decoder, and feature representation. The modules inside the dashed boxes are only used during training.}
    \label{fig1} 
\end{figure}

Furthermore, the widely used time-domain dual-path method \cite{luo2020dual,subakan2021attention,chen2020dual} considers the spatial information of long sequences by dividing the signal into small chunks and processing the intra-chunk and inter-chunk sequences separately. However, it does not consider the sequence's channel (or called feature map) information, which is crucial for speech separation. The reason is that the signal's low and high frequency information are of different importance for speech separation, corresponding to different channels of feature representation in the time domain \cite{luo2019conv,QinZW021}. Low frequency bands usually contain high energies, tonalities, and long-duration sounds, while high frequency bands usually contain low energies, noise, and rapidly decaying sounds \cite{TakahashiM17}. To remedy this deficiency, we use channel attention \cite{woo2018cbam,ParkKSKH22} to re-weight the channel dimension of feature representation to preserve valuable channels and ignore invaluable ones.

The main contributions of this paper are as follows: (\romannumeral1) We propose an end-to-end multi-stage speech separation method to better separate mixtures in noisy and reverberant environments. (\romannumeral2) We propose a triple-path structure to efficiently capture channel information beneficial for speech separation in the time domain. (\romannumeral3) The results show that the proposed training method and model structure can improve speech separation performance with little change in model size.

\section{Problem formulation}

In the noisy and reverberant environment, the monophonic mixture can be described as follows:
\begin{equation}
\small
y(t) = \sum_{i} s_{i}(t) \ast h_{i}(t) + n(t)
\end{equation}
where $t$ is the length index, $s_{i}(t)$, $h_{i}(t)$, and $n(t)$ denote the anechoic speech signal, room impulse response (RIR) of the $i$th speaker, and background noise. $\ast$ represents the convolution operator. The goal is to extract $s_{i}$ from $y$.

\section{multi-stage method}

The target estimated probability $p(s,m,z\mid y)$ can be decomposed using the probability chain rule,
\begin{equation}
\small
p(s,m,z\mid y) = p(z\mid y) \cdot p(m\mid z,y) \cdot p(s\mid m,z,y)
\end{equation}
where $y$ represents the mixture, $z$, $m$, and $s$ represent the denoised mixture, the separated speech with reverberation, and the separated speech. The original difficult problem is decoupled into three sub-problems: speech denoising, separation, and de-reverberation. We call this method multi-stage learning, which can also be regarded as multi-task learning. The advantage is that it can increase the probability of searching for the optimal solution and accelerate the convergence speed of model training by reducing the solution space.

The overall framework of the multi-stage method is shown in Fig. \ref{fig1}. The denoising sub-network converts the encoded feature representation of the mixture into the denoised feature representation in the first stage. The decoder converts it into the denoised mixture $\hat{z}$ during training. The ground truth denoised mixture $z$ is used as the target to calculate the loss $L_1$ to train the denoising sub-network. Similar to the first stage, the separation sub-network uses the denoised feature representation as prior information to predict the separation feature representation of each speaker in the second stage. The decoder converts it into the separated speech with reverberation $\hat{m}$ during training. The denoising sub-network and the separation sub-network are trained by computing the loss $L_2$, where the ground truth separated speech with reverberation $m$ is used as the target. In the third stage, the de-reverberation sub-network predicts the de-reverberation feature representation of each speaker. The decoder converts it into the separated speech $\hat{s}$. The whole model is trained by computing the loss $L_3$ with the ground truth separated speech $s$ as the target. The total loss $L_{\text{total}}$ is:
\begin{equation}
\small
L_{\text{total}}=\alpha L_1+\beta L_2+L_3=\alpha \text{S}(\hat{z},z)+\beta \text{S}(\hat{m},m)+\text{S}(\hat{s},s)
\end{equation}
where the loss function $\text{S}$ is SI-SNR. $\alpha$ and $\beta$ are used to balance the weights of these three loss functions. 

We compare two strategies of denoising, separating, de-reverberating, and denoising, de-reverberating, separating and find that the first strategy yielded better results. The reason may be that if de-reverberation is preceded by separation, the problem is compounded by the fact that the de-reverberation sub-network must adequately compensate for multiple reverberation modes since the multiple sources have different room impulse responses. We do not consider denoising after separation because it is unclear how the network should allocate noise without removing it \cite{maciejewski2020whamr}.

\begin{figure}[t]
    \centering
    \includegraphics[width=0.41\textwidth]{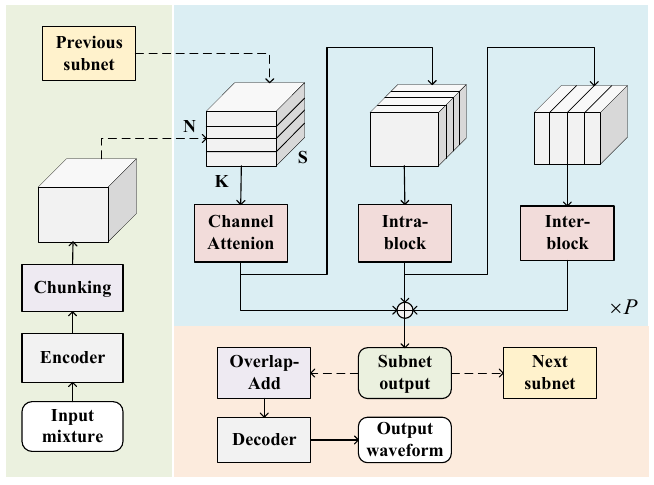}
    \caption{Schematic diagram of the model structure. The triple-path structure in the upper right is shown in blue. Dashed arrows indicate possible inputs or outputs.}
    \label{fig2}
\end{figure}

\section{triple-path structure}
These three sub-problems share similar intrinsic properties of extracting the target signal from the mixture, so the same structure is used to build all sub-networks. We propose a triple-path structure that can model the channel information of audio sequences in the time domain. The overall model structure is shown in Fig. \ref{fig2}. The convolutional encoder encodes the mixture to obtain a $2$-D encoded feature representation, divided into chunks with an overlap factor of $50\%$ in the spatial dimension, making it the input of the first sub-network. 

Channel attention \cite{woo2018cbam} is first used to adaptively predict key information in the channel dimension of feature representation in the triple-path sub-network. Assuming $X$ is the input of channel attention, its weight $\phi(X)$ is calculated as 
\begin{equation}
\small
\phi(X)=\sigma(W_1f(W_0(X_{\text{avg}}))+W_1f(W_0(X_{\text{max}})))
\end{equation}
We apply average and maximum pooling to $X$ to get $X_{\text{avg}}$ and $X_{\text{max}}$. They go through two shared linear layers, where the weights are $W_0$ and $W_1$. $f$ is the ReLU. $\sigma$ is the sigmoid function. The output of channel attention is $\phi(X)\otimes X$, where $\otimes$ represents element-wise multiplication. The pooling operation can extract the general information in the channel dimension and mask the information in the spatial dimension to effectively exploit the correlation between channels \cite{hu2018squeeze}.

Next, the Intra-block and Inter-block model the spatial dimension of sequences globally and locally, which are complementary to the channel attention module. This process is repeated $P$ times. The results of the three paths are added as the output of the sub-network, which is input into the next sub-network. Meanwhile, it is reduced to a $2$-D feature representation by the overlap-add method, which is converted into an audio waveform by a transposed convolutional decoder. The decoders of all sub-networks are parameter-shared. 

All sub-networks predict feature representation instead of masks because the predicted feature representation can be used as prior information for the next sub-network. Additionally, the processing in the previous stages may lead to the loss of clean speech information and introduce artifacts into the predicted feature representation. However, predictive mask-based methods can only deal with additive noise and have difficulty dealing with such interference due to the linear nature of the encoder and decoder \cite{wang2018supervised,chen2020synthesis}.

\section{Experiments}

\subsection{Experimental setup}

The WHAM! \cite{wichern2019wham} and WHAMR! \cite{maciejewski2020whamr} datasets are used to train and evaluate the model. The WHAM! dataset consists of a mixture of two speakers from the WSJ0-2mix dataset \cite{hershey2016deep}, along with unique real ambient noise samples. The WHAMR! dataset is a reverb extension of the WHAM! dataset that further adds artificial reverberation to noisy speech data. SI-SNRi and SDRi are used to evaluate speech separation methods.

To verify the effectiveness of the triple-path structure and the multi-stage method, we add channel attention path to the original dual-path models DPRNN \cite{luo2020dual}, DPTNet \cite{chen2020dual}, and Sepformer \cite{subakan2021attention}, and train them using single-stage and multi-stage methods, respectively. To keep the number of model parameters the same, the number of repetition groups $P$ in each sub-network in the multi-stage method is set to be $1/3$ of the original models since we need three sub-networks with the same number of layers. When experimenting with the WHAM! dataset, $P$ is set to $1/2$ of the original models. We only need denoising and separation sub-networks in this case due to the absence of reverberation in the mixture. 

All sub-networks are jointly trained end-to-end. At the beginning of training, the weights of $L_1$ and $L_2$ are significant to ensure that the learned denoised and separation feature representation really achieves the effect. Then the weight of $L_3$ is gradually increased to stabilize the training process. Expressly, $\alpha$ and $\beta$ are set to start at $1$ and are halved every $20$ epochs of training. For cascaded systems, the denoising, separation, and de-reverberation networks are first trained separately and then jointly optimized end-to-end. The structure of each network is the same as the original dual-path model.

\begin{figure}[t]
    \centering
    \includegraphics[width=0.26\textwidth]{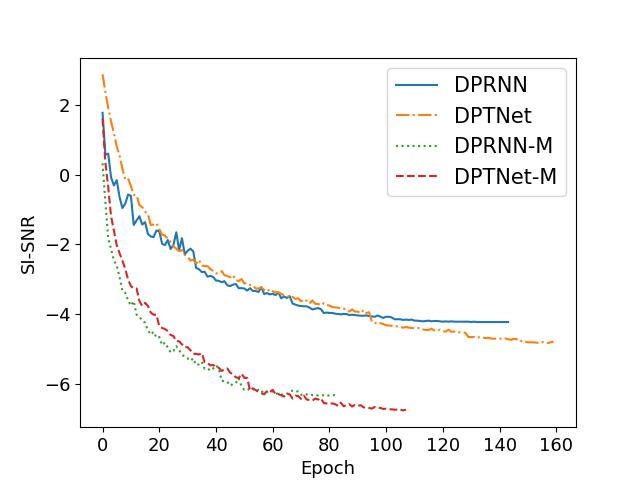}
    \caption{Learning curves of single-stage and multi-stage methods on the WHAMR! dataset. The ordinate represents the value of SI-SNR on the validation set.}
    \label{fig3}
\end{figure}

\begin{table*}[th]
\small
\caption{Model size, SI-SNR and SDR improvements (dB) on the WHAM! and WHAMR! dataset.}
\setlength{\tabcolsep}{5mm}
\label{tab1}
\centering
\begin{tabular}{lrrrrrr}
\hline
\multicolumn{1}{c}{\multirow{2}{*}{\textbf{Method}}} & \multicolumn{3}{c}{\textbf{WHAM!}}                                                                                & \multicolumn{3}{c}{\textbf{WHAMR!}}                                                                               \\
\multicolumn{1}{c}{}                                 & \multicolumn{1}{c}{\textbf{\# Params}} & \multicolumn{1}{c}{\textbf{SI-SNRi}} & \multicolumn{1}{c}{\textbf{SDRi}} & \multicolumn{1}{c}{\textbf{\# Params}} & \multicolumn{1}{c}{\textbf{SI-SNRi}} & \multicolumn{1}{c}{\textbf{SDRi}} \\ \hline
DPRNN \cite{luo2020dual}                                                & 2.6M                                   & 13.9                                 & 14.3                              & 2.6M                                   & 10.6                                 & 10.0                              \\
DPRNN-Ca                                             & 5.2M                                   & 14.5                                 & 14.8                              & 7.8M                                   & 11.7                                 & 10.8                              \\
DPRNN-SRSSN \cite{YaoPCLZ22}                                          & 7.5M                                   & 15.7                                 & 16.1                              & 7.5M                                   & 12.3                                 & 11.4                              \\
\textbf{DPRNN-T}                                     & \textbf{2.7M}                          & \textbf{14.4}                        & \textbf{14.8}                     & \textbf{2.7M}                          & \textbf{11.1}                        & \textbf{10.4}                     \\
\textbf{DPRNN-M}                                     & \textbf{2.7M}                          & \textbf{15.6}                        & \textbf{16.0}                     & \textbf{2.7M}                          & \textbf{12.1}                        & \textbf{11.2}                     \\
\textbf{DPRNN-T-M}                                   & \textbf{2.8M}                          & \textbf{16.2}                        & \textbf{16.6}                     & \textbf{2.8M}                          & \textbf{12.8}                        & \textbf{11.9}                     \\ \hline
DPTNet \cite{chen2020dual}                                               & 2.7M                                   & 14.9                                 & 15.3                              & 2.7M                                   & 11.2                                 & 10.6                              \\
DPTNet-Ca                                            & 5.4M                                   & 15.5                                 & 15.8                              & 8.1M                                   & 11.9                                 & 11.0                              \\
DPTNet-SRSSN \cite{YaoPCLZ22}                                         & 5.7M                                   & 16.1                                 & 16.5                              & 5.7M                                   & 12.3                                 & 11.3                              \\
\textbf{DPTNet-T}                                    & \textbf{2.8M}                          & \textbf{15.3}                        & \textbf{15.7}                     & \textbf{2.8M}                          & \textbf{11.7}                        & \textbf{10.8}                     \\
\textbf{DPTNet-M}                                    & \textbf{2.8M}                          & \textbf{16.0}                        & \textbf{16.3}                     & \textbf{2.8M}                          & \textbf{12.5}                        & \textbf{11.5}                     \\
\textbf{DPTNet-T-M}                                  & \textbf{2.9M}                          & \textbf{16.6}                        & \textbf{16.9}                     & \textbf{2.9M}                          & \textbf{12.9}                        & \textbf{12.3}                     \\ \hline
Sepformer \cite{subakan2021attention}                                            & 25.7M                                  & 15.5                                 & 15.8                              & 25.7M                                  & 13.7                                 & 12.7                              \\
Sepformer + DM \cite{subakan2021attention}                                       & 25.7M                                  & 16.4                                 & 16.7                              & 25.7M                                  & 14.0                                 & 13.0                              \\
Sepformer-Ca                                         & 51.4M                                  & 16.1                                 & 16.5                              & 77.1M                                  & 14.3                                 & 13.5                              \\
\textbf{Sepformer-T}                                 & \textbf{25.8M}                         & \textbf{15.9}                        & \textbf{16.2}                     & \textbf{25.8M}                         & \textbf{14.1}                        & \textbf{13.1}                     \\
\textbf{Sepformer-M}                                 & \textbf{25.8M}                         & \textbf{16.6}                        & \textbf{16.9}                     & \textbf{25.8M}                         & \textbf{14.7}                        & \textbf{14.0}                     \\
\textbf{Sepformer-T-M}                               & \textbf{25.9M}                         & \textbf{16.9}                        & \textbf{17.2}                     & \textbf{25.9M}                         & \textbf{15.2}                        & \textbf{14.4}                     \\ \hline
Wavesplit \cite{ZeghidourG21}                                            & 29M                                    & 15.4                                 & 15.8                              & 29M                                    & 12.0                                 & 11.1                              \\
Wavesplit + DM \cite{ZeghidourG21}                                       & 29M                                    & 16.0                                 & 16.5                              & 29M                                    & 13.2                                 & 12.2                              \\ \hline
\end{tabular}
\end{table*}

\begin{figure}[th]
\centering
\subfigure[]{   
\includegraphics[width=0.11\textwidth]{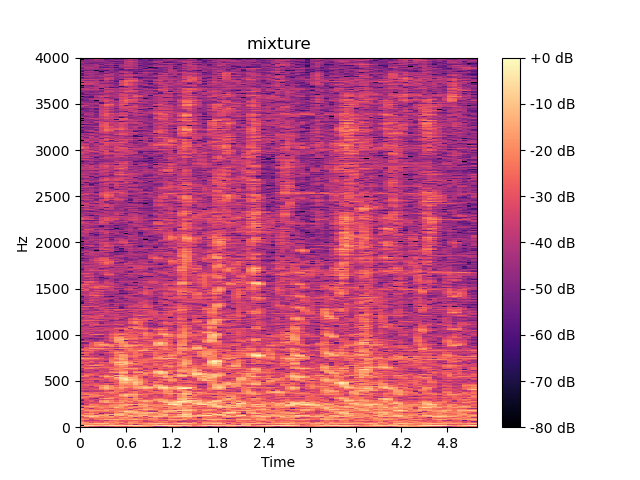}  
}
\subfigure[]{ 
\includegraphics[width=0.11\textwidth]{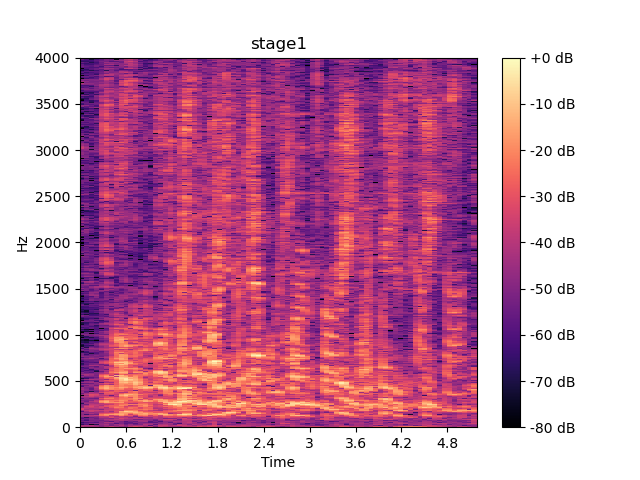}
}
\subfigure[]{   
\includegraphics[width=0.11\textwidth]{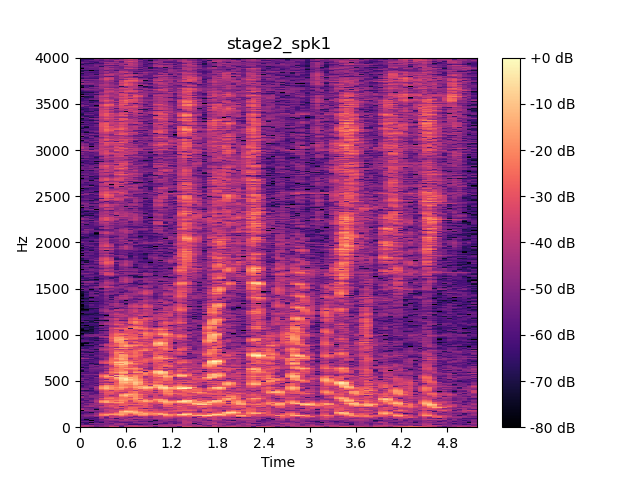}  
}
\subfigure[]{   
\includegraphics[width=0.11\textwidth]{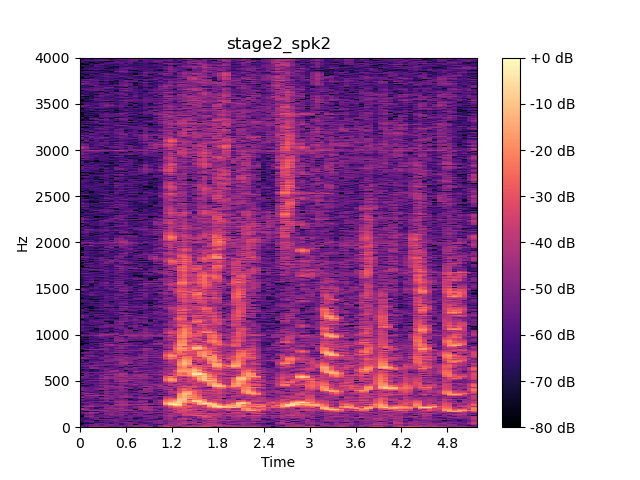}  
}
\subfigure[]{   
\includegraphics[width=0.11\textwidth]{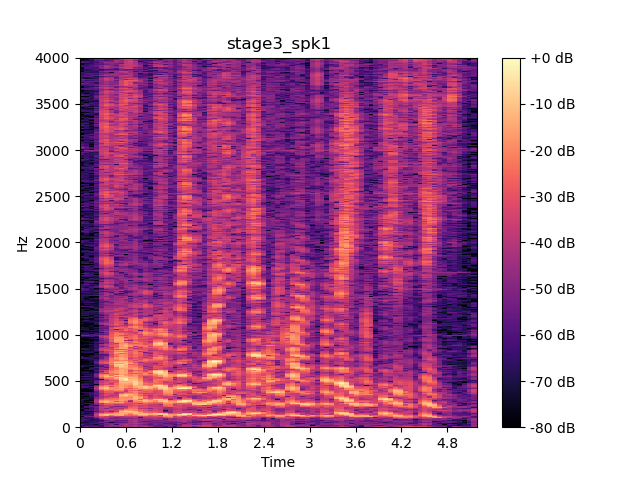}  
}
\subfigure[]{   
\includegraphics[width=0.11\textwidth]{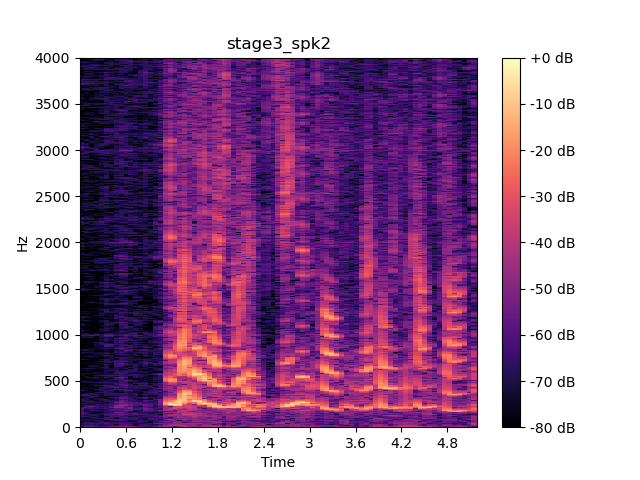}  
}
\subfigure[]{   
\includegraphics[width=0.11\textwidth]{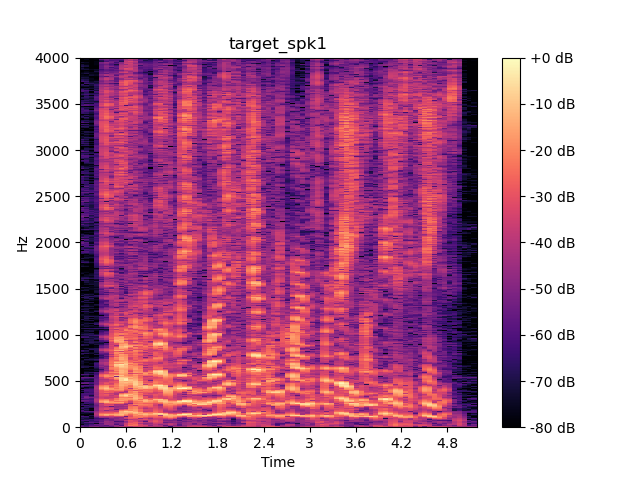}  
}
\subfigure[]{   
\includegraphics[width=0.11\textwidth]{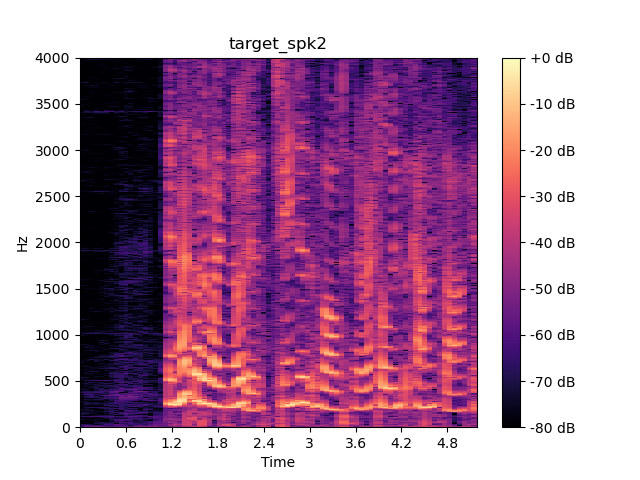}  
}
\caption{Spectrograms of the mixture ((a)), spectrograms of output waveforms of each stage ((b) to (f)), and spectrograms of the target ((g) and (h)).}    
\label{fig4}    
\end{figure}

\subsection{Results and discussion}

The results on the WHAM! and WHAMR! datasets are shown in Table \ref{tab1}. Our methods are shown in bold. -Ca, -T, -M, and DM represent the cascade model, triple-path model, multi-stage learning, and dynamic mixing \cite{ZeghidourG21}. Compared with the original single-stage methods DPRNN, DPTNet, and Sepformer, the multi-stage methods DPRNN-M, DPTNet-M, and Sepformer-M significantly improve the separation effect without substantially increasing the number of parameters and even outperform the cascaded methods DPRNN-Ca, DPTNet-Ca, and Sepformer-Ca with several times the number of parameters. The results suggest that sharing information among multiple tasks leads to better performance than processing each task individually.

Compared with another multi-stage method, the Stepwise-Refining Speech Separation Network (SRSSN) \cite{YaoPCLZ22}, the learning efficiency of our method is higher because our method has more specific objectives in each stage. The results show that our methods DPRNN-M and DPTNet-M can still achieve similar performance when the number of parameters is $1/3$ of DPRNN-SRSSN and $1/2$ of DPTNet-SRSSN. 

It can also be seen from the table that the triple-path models DPRNN-T, DPTNet-T, and Sepformer-T consistently outperform the dual-path models DPRNN, DPTNet, and Sepformer with little increase in the number of model parameters, which verifies the effectiveness of adding channel attention path. Furthermore, the multi-stage triple-path models DPRNN-T-M, DPTNet-T-M, and Sepformer-T-M outperform all baseline models.

We also compare the training convergence speed of the single-stage methods DPRNN, DPTNet, and multi-stage methods DPRNN-M, DPTNet-M, as shown in Fig. \ref{fig3}. The slope of the learning curve of the multi-stage method is larger than that of the single-stage method, and the convergence speed is faster, which verifies that the multi-stage method can improve the probability and speed of searching for the optimal solution by reducing the solution space.

To clarify the state of the multi-stage method at each stage, we visualize the spectrograms of the output of each stage. Fig. \ref{fig4} (a) to (h) represent the spectrograms of the mixture, the outputs of each sub-network, and the target. In the case of two speakers mixing, the denoising sub-network outputs one signal, and the separation and de-reverberation sub-network outputs two signals. Fig. \ref{fig4} (a) and (b) show that the background noise of the mixture is reduced after the denoising stage. Fig. \ref{fig4} (b), (c), and (d) show that the two sources in the mixture are decoupled after the separation stage. Fig. \ref{fig4} (c), (d), (e), and (f) show that the reverberation interference of each source is removed after the de-reverberation stage. The results verify that the multi-stage method achieves the corresponding goal in each stage. Unlike previous speech separation models that operate in a black-box fashion, the proposed framework provides interpretability for each module. An additional benefit of the multi-stage method is that each sub-network can be used for the corresponding task separately. For example, the denoising sub-network can be used for speech enhancement alone after multi-stage training.

\section{Conclusion}
In this study, we derive a novel multi-stage method to solve the problem of speech separation in noisy and reverberant environments via the probabilistic chain rule. Furthermore, we propose a triple-path structure that can adaptively model the channel information of audio sequences in the time domain. Experimental results show that the proposed training method and model structure significantly improve the speech separation models' performance in noisy and reverberant environments without substantially changing the model size.

\vfill\pagebreak

\bibliographystyle{IEEEbib}
\bibliography{strings,refs}

\end{document}